\documentclass{optica-article}
\journal{opticajournal} 

\articletype{Research Article}

\usepackage{float}
\usepackage{booktabs}
\usepackage{graphicx}
\usepackage{caption}
\usepackage{booktabs}
\usepackage{array}
\usepackage{subcaption}

\usepackage{lineno}
\usepackage{dblfloatfix}

\usepackage{amsmath}
\usepackage{graphicx}
\usepackage{cite} 
\setcopyright{}

\begin{document}
\title{Roughness-Limited Performance in Ultra-Low-Loss Lithium Niobate Cavities}
\author{Ali Khalatpour\authormark{*}, Luke Qi\authormark{}, Martin M. Fejer\authormark{}, and Amir H. Safavi-Naeini\authormark{*}}

\address{\authormark{}Department of Applied Physics, Stanford University, Stanford, CA 94305, USA}
\email{\authormark{*}akhalat@stanford.edu, safavi@stanford.edu}
\begin{abstract}
Achieving low optical loss is critical for scaling complex photonic systems. Thin-film lithium niobate (TFLN) offers strong electro-optic and nonlinear properties in a compact platform, making it ideal for quantum and nonlinear optics. While $Q$ factors above $10^7$ have been achieved, they remain below the intrinsic material limit. We present a systematic study of scattering losses due to roughness in TFLN racetrack cavities, isolating contributions from sidewall and interface roughness. Quality factors up to $27 \times 10^6$ are demonstrated in waveguides with widths of $2.2\lambda$ ($\sim3.5\,\mu$m), where interface roughness dominates, and up to $1.2 \times 10^7$ in narrower waveguides $0.8\lambda$ wide ($\sim1.2\,\mu$m), where sidewall roughness is the primary limitation. Our modeling framework, based on 3D wave simulations informed by AFM-measured roughness, is material-independent and broadly applicable across integrated photonic platforms.
\end{abstract}
\section{Introduction}
As integrated photonic systems continue to scale in size and complexity, the demand for material platforms that enable low-loss propagation while supporting both passive and active functionalities becomes increasingly critical. Thin-film lithium niobate (TFLN) presents a compelling solution, offering low optical loss over a broad spectral range—from the visible to the mid-infrared—alongside strong electro-optic and second-order ($\chi^{(2)}$) nonlinearities~\cite{weis1985lithium}. These attributes make TFLN a highly versatile and promising material system for next-generation integrated photonics~\cite{zhu2021integrated}.

While minimizing optical loss has always been central to system optimization, in emerging applications such as optical squeezing~\cite{oposqz,stokowski2023integrated}, single-photon processing~\cite{sund2023high}, and quantum optical computing~\cite{saravi2021lithium}, optical losses directly constrain system realization and yield. As a result, understanding the mechanisms of loss in TFLN photonics and achieving ultra-high quality factors has become essential for the feasibility of the next generation of photonic systems.

The intrinsic absorption of TFLN is estimated to be approximately 0.2~dB/m, corresponding to an absorption-limited quality factor of $\sim 160 \times 10^6$~\cite{ferraro2009ferroelectric, shams2022reduced}. However, fabrication-induced scattering losses often dominate. Lithography processes introduce line-edge roughness which can contribute to optical loss.  Physical etching using high-energy argon ions—currently the most effective method for TFLN—also induces sidewall roughness that increases scattering loss. Ion bombardment can also lead to re-deposition of amorphous lithium niobate on the sidewalls, requiring post-processing steps for removal. Standard cleaning agents such as HF (hydrofluoric acid) or SC-1 (a mixture of ammonium hydroxide and hydrogen peroxide) can further roughen the TFLN surface~\cite{gruenke2024surface}. In addition, the material's sensitivity to high-temperature processes poses further fabrication challenges.

Fabricating high-quality factor devices across both low and high confinement regimes is essential, as performance in these asymptotic limits helps isolate the contributions of sidewall and interface scattering. This not only enables validation of modeling tools but also guides the identification of optimal geometries for complex photonic structures. However, achieving high-$Q$ in the low-mode-volume limit—where optical fields are concentrated near rough surfaces—remains challenging due to extreme sensitivity to fabrication imperfections, despite the appeal of such devices for strong light-matter interactions and nonlinear effects~\cite{park2022high}. In contrast, high-mode-volume devices are less sensitive to fabrication variations and are primarily limited by the intrinsic surface quality of TFLN. Although waveguides much wider than the optical wavelength have demonstrated intrinsic quality factors as high as $29 \times 10^6$~\cite{zhu2024twenty}, their utility is limited by the onset of multimode behavior and reduced nonlinear interaction strength.

While previous studies have examined the role of surface roughness in TFLN propagation loss~\cite{hammer2024estimation,Wolf:18}, a quantitative predictive tool is still lacking. A model that directly predicts propagation loss due to roughness for any given waveguide geometry and surface roughness would be valuable for optimizing TFLN-based devices. Previous approaches based on the Lacey-Payne model (using the equivalent current method~\cite{lacey1990radiation, roberts2022measurements}), originally developed for low-index contrast waveguides, can produce order-of-magnitude errors for high-contrast waveguides ~\cite{barwicz2005three} and are therefore not suitable for TFLN waveguides. In this work, we employ a full-wave simulation method to extract roughness-induced propagation loss from data obtained by atomic force microscopy for the waveguide geometries of interest.
  
\section{Device Architecture Overview}

\begin{figure}[b!]
    \centering
    \includegraphics[width=0.9\textwidth]{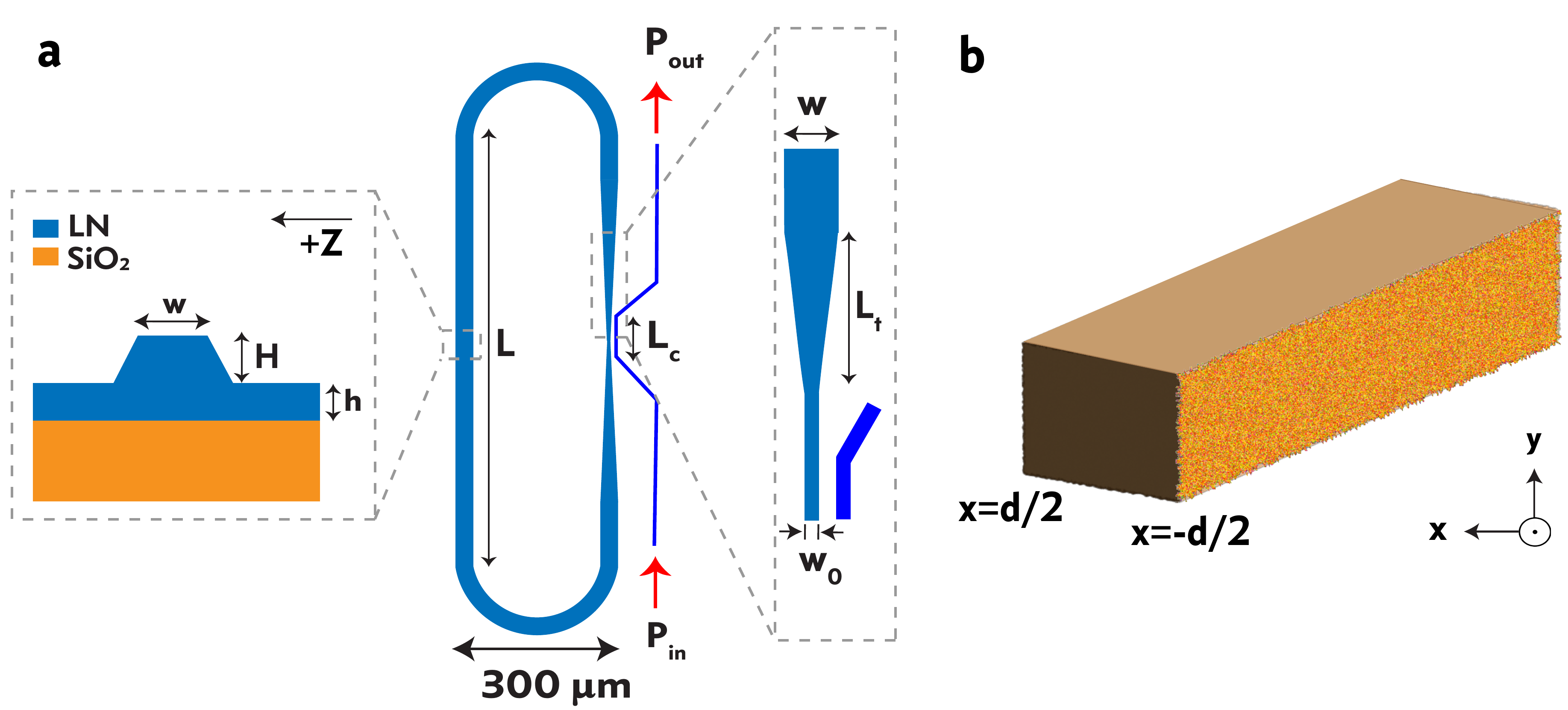} 
    \caption{(a) Geometry of a racetrack and feedline, along with the cross-section of the racetrack resonator characterized by \((w, H, h)\), where \(w\) is the waveguide width, \(H\) is the waveguide height, \(h\) is the slab thickness, \(L\) is the length of the straight section of the racetrack, and \(L_{\text{t}}\) is the taper length (\(200~\mu\)m). 
    (b) Schematic illustration of a waveguide with a rough boundary at \(x = -d/2\).}
    \label{fig:geometry}
\end{figure}

In order to probe the contribution of sidewall roughness to the quality factor, we fabricate racetrack cavities on both X- and Y-cut thin film lithium niobate on oxide wafers (LNOI). The waveguide widths are varied between devices, as wider waveguides are expected to exhibit reduced sidewall scattering loss. To minimize the impact of higher-order modes, we use a narrow and single-mode feedline with a width of \(1.2\,\mu\text{m}\) that is single side-coupled to our cavities. The geometry of the racetracks and feedlines is shown in Figure~\ref{fig:geometry}(a).

In many applications, the thickness of TFLN needs to be adjusted from the thickness provided by the manufacturer (NanoLN in our case). Therefore, to better understand the impact of interface roughness, we prepared two sets of samples, one from X-cut wafers and the other from Y-cut wafers. The X-cut wafers undergo a pre-thinning step using Ar bombardment to remove \(200\,\text{nm}\) of TFLN, whereas the Y-cut wafer surfaces remain at the as-manufactured quality, achieved via chemical mechanical polishing. Since both sets of waveguides have sidewalls oriented in the \(z\)-direction of the crystal, the impact of sidewall roughness is consistent across samples, providing an ideal platform to probe the influence of both top and bottom interface roughness.

\section{Modeling Methodology}
\subsection{Roughness Modeling: Sidewalls}

\label{sec:roughness_modeling}
In this section, we derive an estimate of the power lost by the guided mode due to boundary perturbations along the \(z\)-direction in a waveguide with its cross section in the \((x,y)\) plane. The waveguide boundaries are at \(x=-d/2\) and \(x=d/2\), and we assume that only the boundary at \(x=-d/2\) is perturbed by roughness (see Figure~\ref{fig:geometry}(b)). Suppose that the dielectric tensor takes the form \(\boldsymbol{\epsilon}(x,y,z) = \overline{\boldsymbol{\epsilon}}(x,y) + \delta \boldsymbol{\epsilon}(x,y,z)\), where \(\overline{\boldsymbol{\epsilon}}(x,y)\) describes the ideal \(z\)-independent waveguide and \(\delta \boldsymbol{\epsilon}(x,y,z)\) represents a weak \(z\)-dependent perturbation. The electric field \(\mathbf{E}(x,y,z)\) in a waveguide with a dielectric tensor \(\boldsymbol{\epsilon}(x,y,z)\) satisfies Maxwell's equation
\begin{equation}
\label{curl}
\nabla \times \nabla \times \mathbf{E} + \frac{\omega^2 \mu}{c^2}\,\boldsymbol{\epsilon} \cdot \mathbf{E} = 0.
\end{equation}

We can rewrite~\eqref{curl} as:
\begin{equation}
\label{eq:Maxwell2}
\nabla \times \nabla \times \mathbf{E} 
+ \frac{\omega^2 \mu}{c^2} \bigl(\overline{\boldsymbol{\epsilon}} \cdot \mathbf{E} 
+ \delta \boldsymbol{\epsilon} \cdot \mathbf{E}\bigr) = 0.
\end{equation}

We write the total field as:
\[
\mathbf{E}(x,y,z) = \mathbf{E}_0(x,y)\,e^{i \beta z} + \mathbf{E}_r(x,y,z),
\]
where \(\mathbf{E}_0\) is a known modal field solution of the ideal waveguide and satisfies:
\begin{equation}
\nabla \times \nabla \times \bigl(\mathbf{E}_0(x,y)\,e^{i \beta z}\bigr)
+ \frac{\omega^2 \mu}{c^2}\,\overline{\boldsymbol{\epsilon}} \cdot \mathbf{E}_0(x,y)\,e^{i \beta z} = 0.
\end{equation}

Therefore, substituting \(\mathbf{E}\) into equation~\eqref{eq:Maxwell2} and neglecting second-order correction terms (i.e., \(\delta \boldsymbol{\epsilon} \cdot \mathbf{E}_r\)), we arrive at
\begin{equation}
\label{final}
\nabla \times \nabla \times \mathbf{E}_r 
+ \frac{\omega^2 \mu}{c^2}\,\overline{\boldsymbol{\epsilon}} \cdot \mathbf{E}_r
= -\,\frac{\omega^2 \mu}{c^2}\,\delta \boldsymbol{\epsilon} \cdot \mathbf{E}_0\,e^{i \beta z}.
\end{equation}

Equation~\eqref{final} can be interpreted as an equivalent problem in which the scattered fields, \(\mathbf{E}_r\), are generated by an equivalent current as
\begin{equation}
\label{jeq}
\mathbf{J}_{\mathrm{eq}}(x,y,z) =
-\,i\,\omega\,\delta \boldsymbol{\epsilon}(x,y,z)\cdot\mathbf{E}_0(x,y)\,e^{i \beta z}.
\end{equation}

This derivation corresponds to the so-called Born approximation \cite{jackson1998classical}. The magnetic vector potential corresponding to the equivalent source can be expressed as

\begin{equation}\label{eq:A1}
\mathbf{A}(x,y,x) = \int_{-\infty}^{\infty} \int_{-\infty}^{\infty} \int_{-\infty}^{\infty} 
  \mathbf{J}_{\mathrm{eq}}(x', y', z') \cdot \mathbf{G}(x - x', y - y', z - z') 
  \,dx' \,dy' \,dz',
\end{equation}
where $\mathbf{G}(\mathbf r-\mathbf r')$ is the dyadic Green's function of the unperturbed background, which satisfies the equation $\left[\nabla\times\nabla\times - \tfrac{\omega^2\mu}{c^2}\,\overline{\boldsymbol\epsilon}(x,y)\right]\mathbf{G}(x,y,z)
=\delta(x)\delta(y)\delta(z)$.

To proceed, we further assume that the roughness are perturbations along the sidewall that appear as "bumps" that are uniform along the waveguide height (the $y$ direction). We are therefore ignoring effects of sidewall angle (typically on the order of 10 degress).  Hence, the roughness depends only on $z$, and its variation is described by a small amplitude random function $f(z)$, which is small compared to the cross-sectional dimensions. Therefore, equation \eqref{eq:A1} can be further simplified by using Taylor's expansion as:
\begin{equation}\label{eq:A3}
\mathbf{A} \approx \int_{-\infty}^{\infty} \int_{-\infty}^{\infty} 
  f(z') \,\mathbf{J}_{\mathrm{eq}}(-d/2, y', z') \cdot \mathbf{G}\bigl(x+d/2,\, y - y',\, z - z'\bigr) 
  \,dy' \,dz'.
\end{equation}
This suggests that the source can be expressed as a convolution of a line source with a \(z\)-dependent random function that modulates the phase, as follows:
\[
\mathbf{J}_{\mathrm{eq}}(x,y,z) = \mathbf{J}_L(y) \ast \Bigl\{ e^{i\beta z} \, f(z) \Bigr\}, \quad \text{where} \quad \mathbf{J}_L(y) \equiv \mathbf{J}_{\mathrm{eq}}(-d/2,y,0) \, \delta(x+d/2) \, \delta(z).
\]
To calculate the far-field radiated power of the line source \(\mathbf J_L(x,y,z)\), we can write:
\begin{equation}
\label{eq:Maxwell}
\nabla \times \nabla \times \mathbf{A}_L 
+ \frac{\omega^2 \mu}{c^2}\, \overline{\boldsymbol{\epsilon}} \cdot \mathbf{A}_L 
= \mu_0 \mathbf J_L.
\end{equation}
We start with the magnetic vector potential \( \mathbf A_L \) associated with a line source represented by the current density \( J_L \). Magnetic potential \( A(x, y, z) \) can be expressed as a convolution integral:
\begin{equation}
\mathbf A(x, y, z) = \int \int \int \mathbf A_L(x - x', y - y',
z - z') f(z') e^{i \beta z'} \, dx' \, dy' \, dz'.
\label{eq:vector_potential}
\end{equation}

The radiation modes in the Fourier domain lie on a surface of a sphere with radius \(k = n_{\text{clad}} k_0\). Thus, the ensemble average of the intensity of radiated field in the far field is given by: 
\begin{equation}
\langle P_{\text{rad}} \rangle = \frac{1}{(2 \pi)^2}\frac{1}{2\eta} \int_{|\mathbf k|=n_{\text{clad}}k_0} \langle |\hat{\mathbf E}(\mathbf k)|^2 \rangle\, k^2\, d\Omega,
\label{p_far}
\end{equation}
Here we used \(\langle \rangle\) for ensemble averaging and \(\eta\) for the wave impedance in the cladding material, and the $\hat{\mathbf E}(\mathbf k)$ is the Fourier transform of the electric field.
By employing the properties of the convolution operator in the Fourier domain, and using the far-field approximation with the relationship \(E = -i \omega A\), the ensemble-averaged Fourier intensity is given by:
\begin{equation}
\langle |\hat{\mathbf E}(\mathbf k)|^2 \rangle = L |\hat{\mathbf E}_L(\mathbf k)|^2\,\hat{R}(\beta - k_z),
\label{eq:FourierIntensity}
\end{equation}
where $\hat{R}(k_z) \equiv \langle |\hat{f}(k_z)|^2 \rangle/L$ is the power spectral density of $f(z)$, and $\hat{\mathbf E}_L(\mathbf k)$ represents the Fourier transform components of the far field from the line source.

Therefore, we can write:
\begin{equation}
\frac{\langle P_{\rm rad}\rangle}{L}
=\int_{|\mathbf k|=n_{\rm clad}k_0}
  \frac{k^2}{2\eta}\,\bigl|\hat{\mathbf E}_L(\mathbf k)\bigr|^2\,
  \hat R\bigl(\beta - k_z\bigr)\,d\Omega.
\label{p_far_result}
\end{equation}
Defining \(S_L(\theta,\phi)\equiv\frac{1}{(2 \pi)^2}\frac{k^2}{2\eta}\left|\hat{\mathbf E}_L(\mathbf k)\right|^2\) in \eqref{p_far_result} the far-field scattering power per unit length can be approximated as
\begin{equation}
\frac{\langle P_{\text{rad}} \rangle}{L} = \int_0^{2\pi} \int_0^{\pi} S_L(\theta,\phi)\,\hat{R}\Bigl(n_{\text{eff}} k_0 - n_{\text{clad}} k_0 \cos \theta\Bigr)\,\sin \theta\,d\theta\,d\phi,
\label{eq:final_expression}
\end{equation}
This yields the final expression for the scattering (or absorption) coefficient $\alpha_s$, which is defined by the equation $P_\text{rad}/L=\alpha_s P_G$, giving us:
\begin{equation}
\alpha_s = \frac{1}{P_G}\displaystyle \int_0^{2\pi} \int_0^{\pi} S_L(\theta,\phi)\,\hat{R}\Bigl(n_{\text{eff}} k_0 - n_{\text{clad}} k_0 \cos \theta\Bigr)\,\sin \theta\,d\theta\,d\phi,
\label{eq:scattering_coefficient}
\end{equation}
where \(P_G\) is the guided mode power (i.e., the integral of the \(z\)-component of the Poynting vector across the cross-section). If the roughness profiles for the left and right sidewall are uncorrelated, then each sidewall contributes independently to the total scattering coefficient, so the total is $\alpha_s^{\text{tot}} = \alpha_s^{(-d/2)}+\alpha_s^{(+d/2)}$. In the common special case where the sidewalls have identical roughness statistics, this simply gives us $\alpha_s^{\text{tot}}=2 \alpha_s^{\text{single sidewall}}$, giving an equation which is identical to that used in \cite{barwicz2005three}.

\subsection{Roughness Modeling: Top and Bottom Surfaces}

\label{sec:interface-roughness}
In ridge-waveguide geometries, scattering that arises from roughness at the top and bottom interfaces can be described by a two-dimensional random function. Let \( f(x,z) \) denote isotropic height variations at a rough interface at \( y=0 \). We assume \( f(x,z) \) is a zero-mean, stationary random function with power spectral density (PSD) \( \hat{R}_f(k_x,k_z) \). Following the same approach used in \eqref{eq:A3}, we can express the scattered magnetic vector potential \( \mathbf{A} \) in the form:
\begin{equation}
\label{eq:A4}
\mathbf{A}(x,y,z)
\approx
\int_{-\infty}^{\infty} \int_{-\infty}^{\infty} 
f(x',z')\,
\mathbf{J}_{\mathrm{eq}}(x',0,z')\,
\mathbf{G}\bigl(x - x',\,y,\,z - z'\bigr)\,
\mathrm{d}x'\,\mathrm{d}z',
\end{equation}
where \( \mathbf{J}_{\mathrm{eq}}(x,0,z) \) is the equivalent surface current of the fundamental mode. Here we approximate the profile of the fundamental TE mode of a ridge waveguide as a cosine with length scale $d'$ in the transverse direction. The surface current is given this cosine term modulating the roughness function \( f(x,z) \) over a waveguide width $d$ as:
\begin{equation}
\label{eq:g_def}
J(x,z)=f(x,z)\,\cos\left(\frac{\pi x}{d'}\right)\,\mathrm{rect}\left(\frac{x}{d}\right)\,e^{i \beta z}.
\end{equation}
Here \( d' \) is the extent of the mode (always larger than width, \( d \), though \( d' \approx d \) for wide waveguides), and \( \mathrm{rect}(u) = 1 \) for \( |u| \le \tfrac{1}{2} \) and zero otherwise. To find \(\hat{R}_J(k_x,k_z) = \langle |\hat{J}(k_x,k_z)|^2 \rangle\), we start by defining \( g(x) \) where \( J(x,z) = f(x,z)e^{i \beta z}\,g(x) \). Then:

\begin{equation}
\hat{J}(k_x,k_z)
=
\frac{1}{2\pi}\int_{-\infty}^{\infty}\hat{f}(k'_x,k_z-\beta)\,\hat{g}(k_x - k'_x)\,dk'_x.
\label{eq:J_convolution}
\end{equation}

The power spectral density (PSD) $\hat{R}_J(k_x,k_z)$ of the random function $J(x,z)$ is defined as the ensemble average of the squared magnitude of its Fourier transform:

\begin{equation}
\hat{R}_J(k_x,k_z) 
\;=\; 
\left\langle |\hat{J}(k_x,k_z)|^2 \right\rangle.
\end{equation}

\begin{equation}
\hat{R}_J(k_x,k_z)
=
\frac{1}{(2\pi)^2}
\int_{-\infty}^{\infty}
\hat{R}_f(k'_x,k_z-\beta)\,\bigl|\hat{g}(k_x - k'_x)\bigr|^2\,dk'_x.
\label{eq:RJ_step4_final}
\end{equation}

The Fourier transform of g(x) can be expressed as: 
\begin{equation}
\label{sincs}
\hat{g}(k_x)
\;=\;
\frac{d}{2}\Bigl[
 \mathrm{sinc}\Bigl(\tfrac{d}{2}\bigl(k_x - \tfrac{\pi}{d'}\bigr)\Bigr)
 +
 \mathrm{sinc}\Bigl(\tfrac{d}{2}\bigl(k_x + \tfrac{\pi}{d'}\bigr)\Bigr)
\Bigr].
\end{equation}
In \eqref{sincs} we have \(d' > d\), so the two \(\mathrm{sinc}\) lobes overlap slightly.  
The peaks are separated by \(2\pi/d'\), whereas each main lobe has a width of order \(2\pi/d\);  
hence the lobes are not completely isolated, and their overlap introduces a small cross term.  
 Only in the limit \(d' \to d\) (i.e., for very wide waveguides) does the peak of one \(\mathrm{sinc}\) nearly align with the first zero of the other, thereby suppressing interference. Throughout the remainder of the analysis we neglect this cross term and, for simplicity, replace each
\(|\mathrm{sinc}|^{2}\) factor by a delta function of equal area. This yields:
\begin{equation}
\hat{R}_J(k_x,k_z)
\;\approx\;
\frac{1}{2\pi}\,\frac{d}{4}
\left[
\hat{R}_f\left(k_x - \tfrac{\pi}{d'},\,k_z - \beta\right)
+
\hat{R}_f\left(k_x + \tfrac{\pi}{d'},\,k_z - \beta\right)
\right].
\label{eq:rg_final}
\end{equation}
The scattering loss can then be calculated similarly as:
\begin{equation}
\alpha_s = \frac{1}{P_G}{\displaystyle \int_0^{2\pi} \int_0^{\pi} S_p(\theta,\phi)\,\hat{R}_J(k_x,k_z)\,\sin \theta\,d\theta\,d\phi},
\label{eq:scattering_coefficient2}
\end{equation}
where \( S_p(\theta,\phi) \) is the power radiated per unit solid angle for a delta-function source defined as \( \mathbf{J}_p = \mathbf{J}_0 \, \delta(x) \, \delta(y) \, \delta(z) \). Here, \( \mathbf{J}_0 \) corresponds to the equivalent current at \( \mathbf{E}_0 \) at \( x = y = z = 0 \).

\noindent

 Since the radiation intensity of a subwavelength radiator scales with its physical length squared (i.e., \( \propto L^2 \) for \( L \ll \lambda \)), equation \eqref{eq:scattering_coefficient2} can be rewritten in the following form:

\begin{equation}
\alpha_s = \left( \frac{d}{L_c} \right) \left( \frac{1}{L_c} \right) 
\frac{\displaystyle \int_0^{2\pi} \int_0^{\pi} 
S_{\text{patch}}(\theta,\phi)\,
\hat{R}_J(k_x,k_z)/(dL_c^2)
\,\sin \theta\,d\theta\,d\phi}{P_G},
\label{eq:scattering_coefficient3}
\end{equation}

\noindent
where \( S_{\text{patch}}(\theta,\phi) \) is the far-field radiation from a coherent patch of size \( L_c \times L_c \) with uniform current \( J_0 \). This form highlights that the prefactor represents the number of statistically independent scattering patches, interpreting roughness-induced loss as the sum of their uncorrelated radiation.

\subsection{From propagation loss to intrinsic quality factor of a resonator}

Throughout the remainder of the paper we quote and compare \emph{quality factors}, because
\(Q\) is what we extract directly from the optical linewidth.  Here we connect these measured quantities to the scattering–loss coefficient \(\alpha_s\) that
comes from the roughness model. Our on-chip power decays as \(P(z)=P_0 e^{-\alpha_s z}\) from which we obtain
\[ Q_{\text{i}}
   =\frac{\omega_0}{v_g\,\alpha}.
\]
Generally, the propagation loss extracted from the scattering model we use does not translate uniquely into a single quality factor. This is because the scattering may be into higher-order modes of the waveguide. These modes may themselves resonate inside the cavity. The net effect of this scattering can then be enhanced or suppressed depending on the detuning of the mode of interest from the higher-order resonances of the cavity.  In our case, we are assuming that the sidewall and interface loss are effectively irreversible. This may be a good assumption if, for example, there is enhanced out-coupling, scattering, and bend loss in higher-order modes. Then, the energy scattered into a higher mode can be assumed to be effectively lost, and so these feedback effects are not taken into account. In this work, we use the simple relation between $Q$ and $\alpha$ described above and ignore these effects.
 
\section{Measurements and Discussion}

As shown in our theoretical work, our approach requires the far-field radiation pattern of a small patch or a line source, along with the roughness parameters, to predict the scattering loss and infer the quality factor of the resonator. The details of the 3D simulation, which we performed using a 3D FDTD solver (\texttt{Tidy3D}) are provided in the Supplemental Material. In addition, due to the discontinuity of electric fields at the sidewall for the fundamental TE mode, we used the electrostatic limit and polarizability matrix to extract the equivalent current~\cite{johnson2005roughness}, as detailed in our Supplemental Material. In the following section, we proceed with the extraction of roughness parameters.

\begin{figure}[h!]
    \centering
    \includegraphics[width=.95\linewidth]{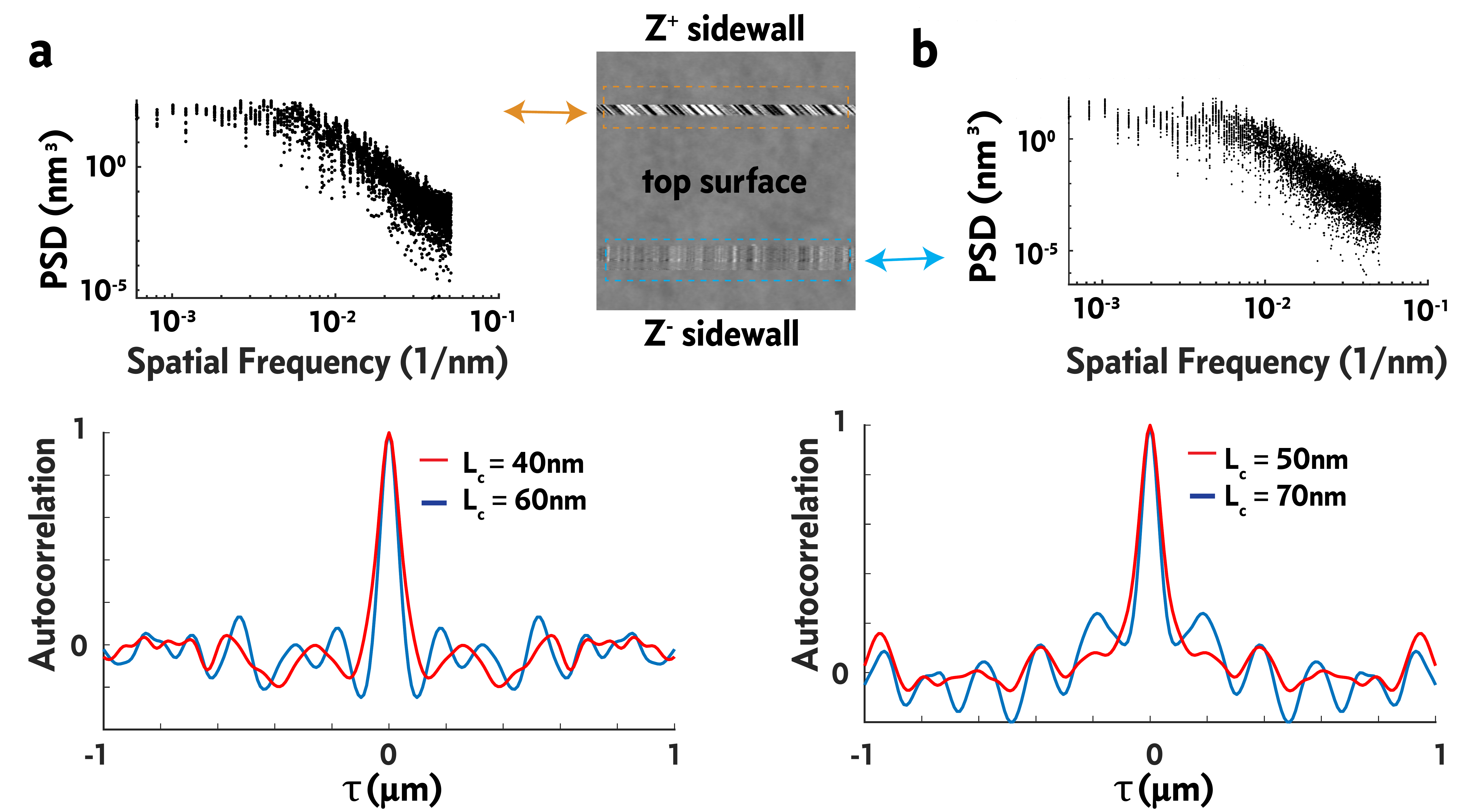}
    \caption{\small 
    Autocorrelation function of the roughness obtained from AFM measurements 
    on the \(Z^+\) and \(Z^-\) sides of a lithium niobate waveguide with width 
    \(w = 1.2\)\,\(\mu\)m. The two lines mark the range of the correlation length \(L_c\). 
    The insets show (i) a 3D AFM image indicating line-like roughness features 
    propagating through the sidewall and (ii) the sidewall PSD with a corner frequency, 
    validating the scan length and resolution.
    }
    \label{fig:AFM}
\end{figure}

\subsection{Evaluating the Roughness by Atomic Force Microscopy}

We performed atomic force microscopy (AFM) in tapping mode using an Asylum Research AFM and a Bruker TESPA-V2-SS cantilever tip. To extract the roughness parameters, we conducted two-dimensional scans over areas of 2--5~\(\mu\)m \(\times\) 5~\(\mu\)m, covering both the sidewall and the top surface of the waveguides.
Figure~\ref{fig:AFM} presents the autocorrelation of the \(Z^+\) and \(Z^-\) surfaces extracted from AFM scans of a device with width \(w=1.2\)\,\(\mu\)m. The two lines represent the range of the correlation length \(L_c\). The insets show a top-view AFM scan and the sidewall power spectral density (PSD). Although the sidewall PSD does not follow a purely Lorentzian profile, it exhibits a slowly varying region with a kink at higher frequencies (corresponding to \(1/L_c\)), confirming that the scan length and resolution were sufficient to capture the corner frequency. We approximated \(L_c\) by analyzing the autocorrelation of the scan lines. The 3D AFM perspective (see Fig.~\ref{fig:AFM}) indicates that the roughness appears as line-like features that propagate along the sidewall, supporting a modeling approach that neglects variations along the waveguide height.

Table~\ref{tab:roughness} summarizes the roughness parameters (\(\sigma\)) and correlation lengths (\(L_c\)) for devices with widths of 1.2, 2.5, and 3.5\,\(\mu\)m. The \(Z^+\) surface shows significantly higher roughness than the \(Z^-\) surface. Wider devices exhibit increased sidewall roughness, likely due to overexposure of the lithography mask. This can be mitigated by optimizing the exposure dose for larger waveguide widths. Due to the extremely smooth surface, tapping mode is not appropriate for measuring the roughness of the top surface. We performed a ScanAsyst scan (peak force mode) for those surfaces, and a detailed section is presented in the supplemental file.

\begin{table}[h!]
\centering
\caption{Roughness parameters \(\sigma\) and correlation length \(L_c\) for different sidewall widths.}
\label{tab:roughness}
\renewcommand{\arraystretch}{1}
\setlength{\tabcolsep}{4pt} 
\begin{tabular}{
    >{\centering\arraybackslash}p{2.4cm}
    >{\centering\arraybackslash}p{1.5cm}
    >{\centering\arraybackslash}p{1.5cm}
    >{\centering\arraybackslash}p{1.5cm}
    >{\centering\arraybackslash}p{2.2cm}}
\toprule
 & \multicolumn{3}{c}{\textbf{Width} (\(\mu\)m)} & \\
\cmidrule(lr){2-4}
 & \textbf{1.2} & \textbf{2.5} & \textbf{3.5} & \(\mathbf{L_c}\) (nm) \\
\midrule
\(\sigma_{z^+}\) (nm)  & \(1.4\text{–}1.6\) & \(2.2\text{–}2.4\) & \(2.2\text{–}2.45\) & \(40\text{–}60\) \\
\(\sigma_{z^-}\) (nm)  & \(0.4\text{–}0.5\) & \(0.4\text{–}0.5\) & \(0.4\text{–}0.5\) & \(45\text{–}65\) \\
\bottomrule
\end{tabular}
\end{table}

\subsection{Quality Factor Measurements}

\begin{figure}[h!]
    \centering
    \includegraphics[width=.75\linewidth]{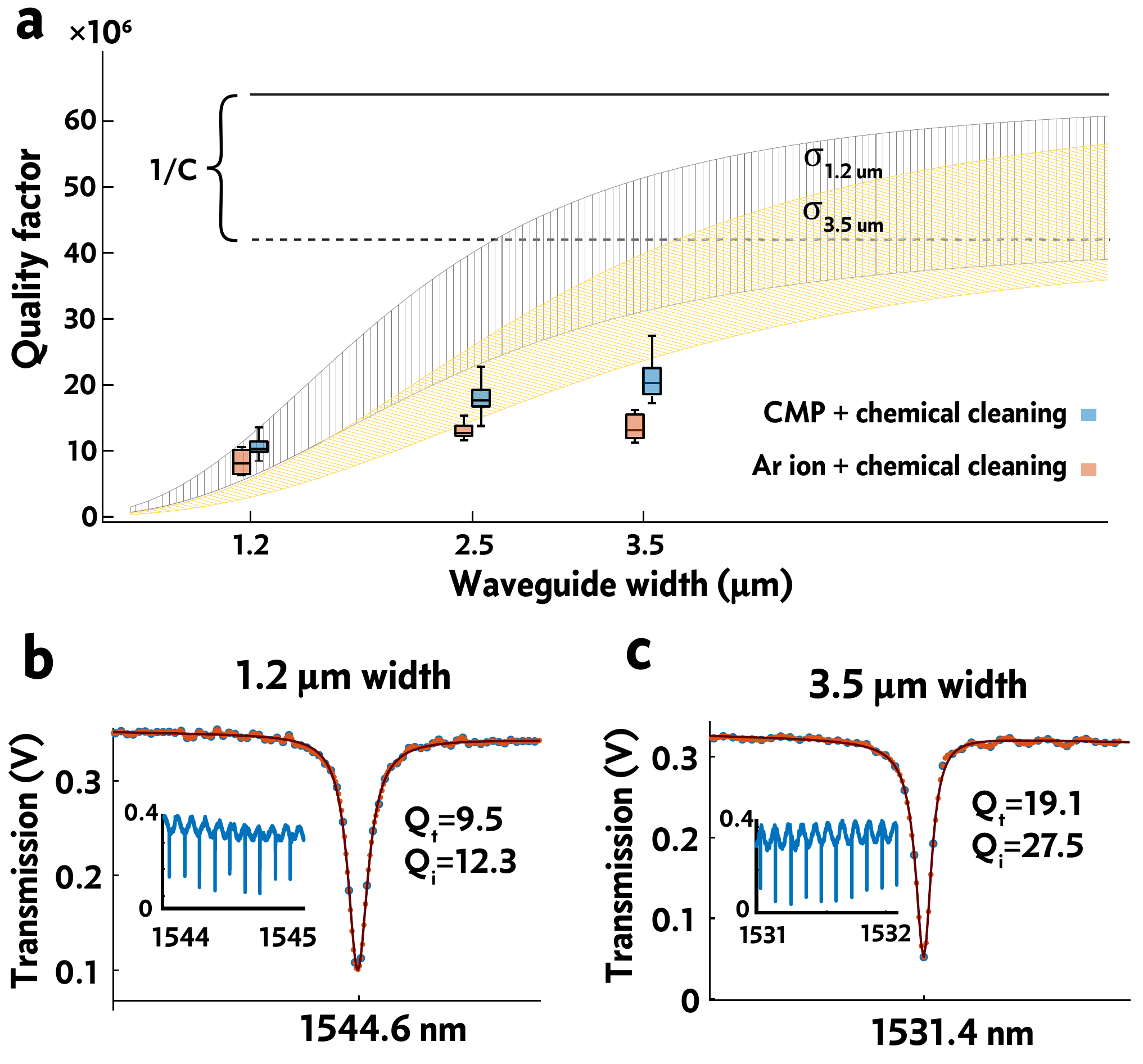}
    \caption{\small 
    (a) Comparison of experimentally measured \(Q\) values (boxplots) and simulated \(Q\) for waveguides with \(H = 300\,\text{nm}\), \(h = 200\,\text{nm}\), and widths \(w = 1.2\,\mu\text{m}\), \(2.5\,\mu\text{m}\), and \(3.5\,\mu\text{m}\). The black lines indicate the estimated bounds for the constant term \(1/C\) [see Eqs.~\eqref{eq:Q_combined}]. The gray shaded region shows the total expected loss including intrinsic, interface, and sidewall contributions for the $1.2\ \mu$m waveguide roughness, and the area is the uncertainty in the AFM measurement in Table~\ref{tab:roughness}. Similarly the yellow region corresponds to the $3.5\ \mu$m waveguide roughness.
    (b) Transmission response and corresponding measured \(Q\) for the single-mode \(1.2\,\mu\text{m}\) device. 
    (c) Transmission response and measured \(Q\) for the single-mode \(3.5\,\mu\text{m}\) device.}
    \label{fig:Experimental}
\end{figure}

The experimental data, along with comparisons to simulated values, are shown in Figure~\ref{fig:Experimental}. Details concerning the extraction of quality factors from measured transmission spectra, as well as the statistical parameters characterizing the top interface, are provided in the Supplemental Material. Here, we decompose the quality factor into a width-dependent contribution from the sidewall and a nearly constant term, according to
\begin{equation}
\frac{1}{Q} = \frac{1}{Q_{\text{sidewall}}} + \frac{1}{Q_{\text{top}}} + \frac{1}{Q_{\text{bottom}}} + \frac{1}{Q_{\text{absorption}}}=\frac{1}{Q_{\text{sidewall}}}+\frac{1}{C}\text{},
\label{eq:Q_combined}
\end{equation}
Our scattering model analysis which combines the AFM-derived roughness parameters for the top, bottom, and independently measured bulk absorption~\cite{shams2022reduced} lead us to a best estimate for \(1/C\) of \(42\text{–}64 \times 10^{6}\) (see Supplementary Materials). 
We draw two observations from the data. First, both Ar-trimmed and CMP-trimmed surfaces yield similar quality factors for narrow waveguides (on the order of $10^7$), in agreement with our model. This confirms that in narrow devices, losses are dominated by roughness of the sidewall. In contrast, CMP-polished surfaces exhibit higher average $Q$ values in larger devices, where the top and bottom interface losses become more significant. The superior surface quality achieved through CMP, as measured in this work and reported in the literature~\cite{gruenke2024surface}, supports this conclusion. It should be noted that the final surface quality of these waveguides is ultimately influenced by subsequent chemical processing. However, as shown in the Supplemental Material, CMP-polished surfaces still provide interfaces superior to those of Ar-trimmed samples, even after identical chemical treatments.

A similar observation was reported in Ref.~\cite{zhu2024twenty}, where $Q \approx 2.9 \times 10^7$ was achieved only in very wide waveguides ($w = 4.5\,\mu\mathrm{m}$, $L = 10\,\mathrm{mm}$). By comparison, our results demonstrate that comparable $Q$ values can be attained in narrower waveguides, which are more attractive due to their larger nonlinear and electro-optic coupling.

In general, these experiments show that for narrow waveguide resonators, losses are dominated by sidewall roughness, which scales as $\sigma^2$. Minimizing edge roughness through improved fabrication is crucial for achieving ultra-high-$Q$ devices. In this regime, our model predicts the observed $Q$ values, and points out at what level of performance the top and bottom surfaces will begin to dominate quality factors.

\section{Concluding remarks}

A comprehensive understanding of loss mechanisms in TFLN waveguides is essential for continued progress in the field of nonlinear and quantum photonics. Our results provide a quantitative framework for estimating the relative contributions of scattering mechanisms. The methodology we have demonstrated is material-agnostic and more broadly applicable across the integrated photonics community. Importantly, this study informs us of which aspects of the fabrication process are the most fruitful to focus on for rapid progress. By systematically identifying and addressing these losses, we believe that realizing the material limits of TFLN in real devices is possible, and will lead to unprecedented device performance and new capabilities.

\section{Acknlowledgements}

This work was supported by the U.S. government through the Defense Advanced Research Projects Agency (DARPA) INSPIRED program, the National Science Foundation NSF-SNSF MOLINO project No. ECCS-2402483, and the US Department of Energy through grant no. DE-AC02-76SF00515 and via the Q-NEXT Center. Device fabrication was performed at the Stanford Nano Shared Facilities (SNSF) and the Stanford Nanofabrication Facility (SNF), supported by the National Science Foundation under award ECCS-2026822. This material is based upon work supported by the National Science Foundation Graduate Research Fellowship Program under Grant No. (DGE-1656518). L.Q. also gratefully acknowledges support from the Shoucheng Zhang Graduate Fellowship Program. The authors appreciate financial and technical support from NTT Research. 

\bibliography{main.bib}
\pagebreak
\section{Supplemental Material}

\subsection{Crystal Axis Dependent Etch in LN} 

As shown in the main text, the impact of sidewall roughness can be reduced
by increasing the waveguide width. We also observed that for narrower devices
(\(w \approx 1.2\,\mu\text{m}\)), the dominant scattering arises from the
sidewalls, whereas for wider devices (\(w \approx 2.5\,\mu\text{m}\)), 
the primary contribution to loss comes from the top and bottom interfaces. 

Periodically poled lithium niobate (PPLN) waveguides 
(e.g., \cite{batchko1999backswitch,bordui1991compositional,Wang:18}) 
are of interest for many applications. In PPLN, the optical axis of LN 
(\(+Z\)) is periodically flipped by applying a strong electric field of 
approximately \(3\,\text{V}/\mu\text{m}\). This poling process can artificially increase the crystal-dependent, fabrication-induced roughness, 
because the chemical processing used here has different etch rates and
roughness profiles on the \(+Z\) and \(-Z\) faces of LN.

\begin{figure}[h!]
    \centering
    \includegraphics[width=0.7\linewidth]{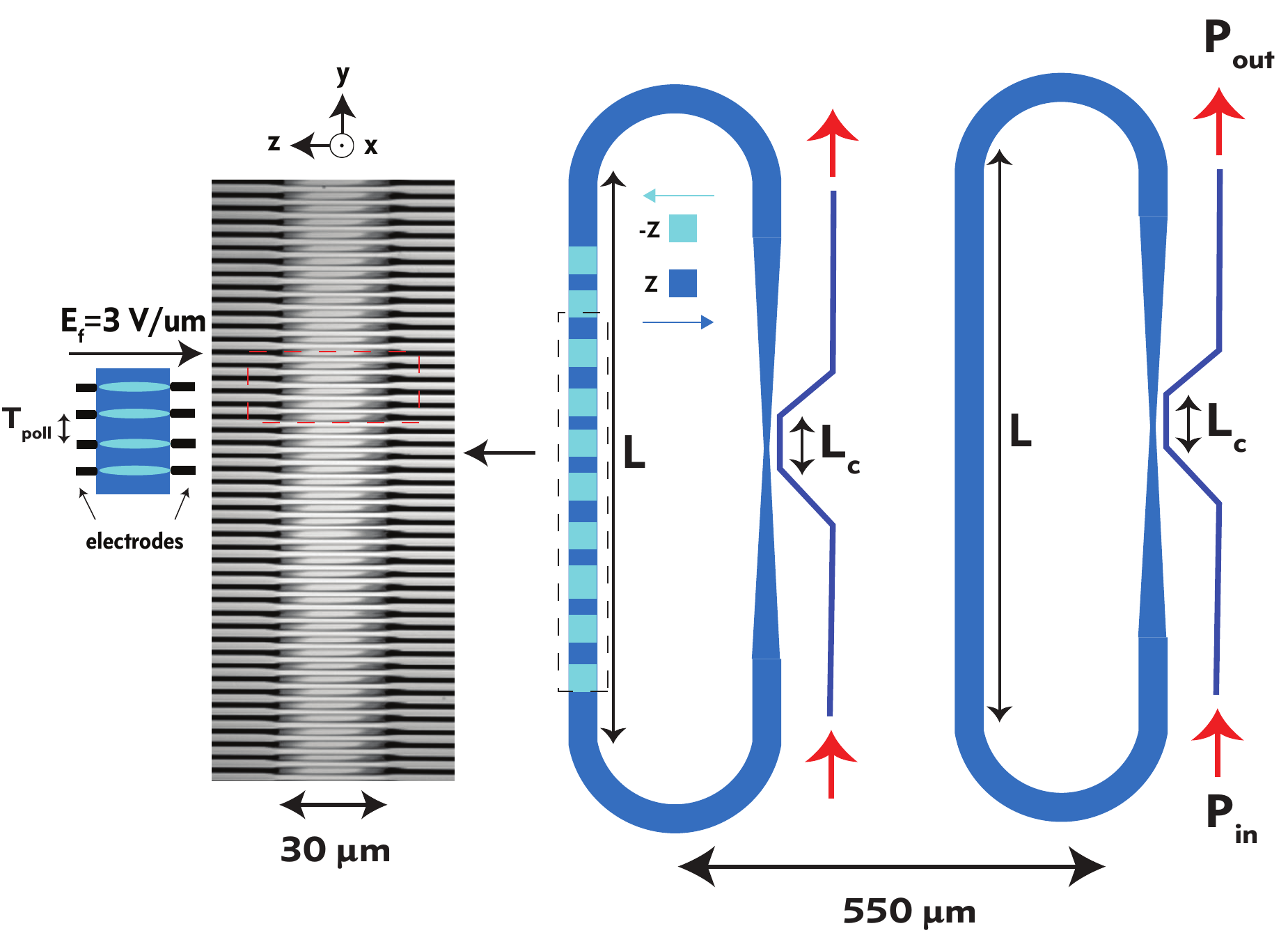}
    \caption{\small 
        Schematic of our experimental setup, where the false-colored 
        section represents the poled region in a \(2\,\text{mm}\) racetrack. 
        An SHG microscope image of the domain formation 
        with \(T_{\text{pol}} = 3.7\,\mu\text{m}\) is shown alongside 
        a schematic for clarity. An electric field is applied to periodically 
        change the crystal optical axis (\(z\)). The darker lines represent 
        the poling fingers (aluminum), and the brighter areas show the depth 
        of the poled region in the slab. The poling region is \(30\,\mu\text{m}\) 
        wide and extends uniformly across the waveguide width. Alignment marks 
        ensure the waveguide is centered in the poled slab.}
    \label{fig:poll-setup}
\end{figure}

Since the scattering loss of LN in narrow waveguides is dominated by
sidewall roughness, we expect that any additional roughness from poling
will lead to pronounced loss in narrower devices, while wider devices
should be comparatively unaffected. To explore this effect, we constructed a total length racetrack resonator \(2.5\,\text{mm}\) in which one arm
(straight section) was positioned, representing approximately 40\% of the total
cavity length. We tested two devices on the same chip, separated by
\(550\,\mu\text{m}\) to minimise processing variations. One device had 
a waveguide width of \(w = 1.2\,\mu\text{m}\), and another was tapered 
to \(w = 2.5\,\mu\text{m}\). We poled one device, using the neighboring 
(non-poled) device as a baseline. Aside from this difference, all fabrication 
steps were identical—including exposure to chemicals, solvents, and crucial 
sidewall cleaning processes. Any observed differences are therefore 
attributed directly to these crystal-axis-dependent fabrication processes. 
The experimental setup is illustrated in Figure~\ref{fig:poll-setup}. 

\begin{figure}[h!]
    \centering
    \includegraphics[width=0.9\linewidth]{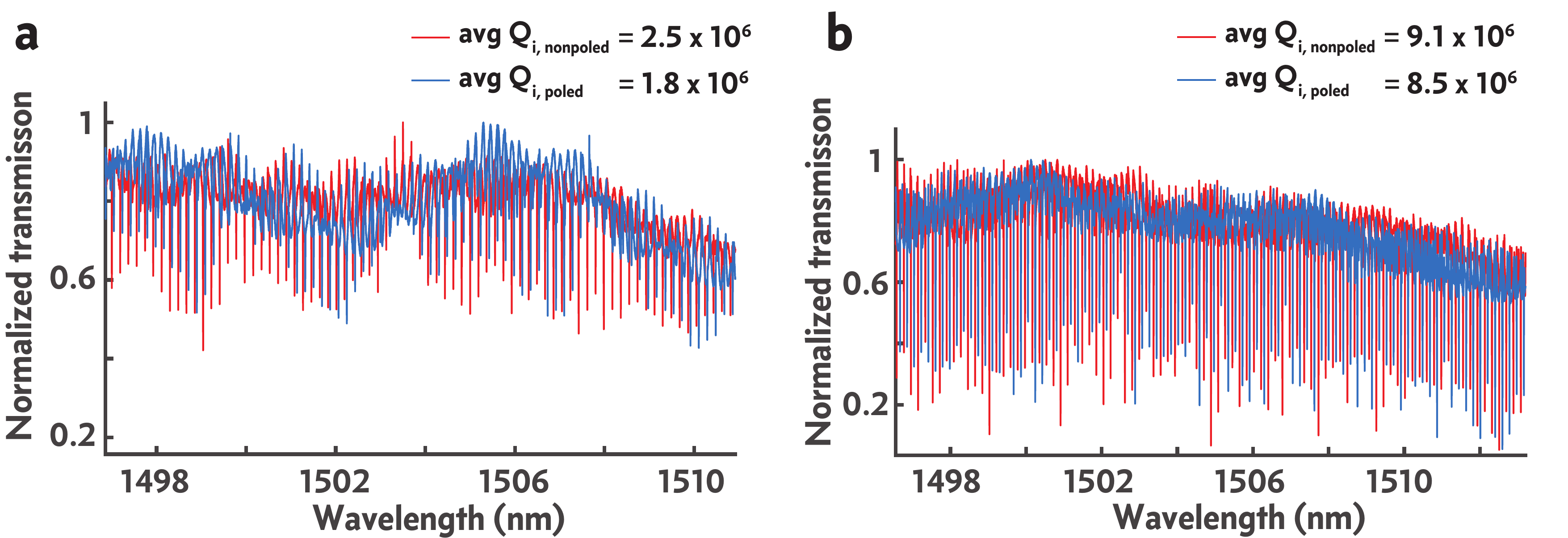}
    \caption{\small 
       Quality factors of racetrack resonators with \(w = 1.2\,\mu\text{m}\) 
       and \(w = 2.5\,\mu\text{m}\) before and after poling. The narrower 
       device (left) exhibits more than a 30\% degradation, whereas the 
       wider device (right) experiences less than a 7\% reduction in 
       quality factor. This result confirms that wider devices are 
       predominantly affected by top and bottom interfaces, as derived 
       in the main text.}
    \label{fig:poll}
\end{figure}

As shown in Figure~\ref{fig:poll}, the narrow device experiences a 
degradation of more than 30\%, whereas for the wider device the degradation 
is less than 7\%, consistent with our expectations. A more quantitative 
analysis of the poled devices will be presented in a separate study.

\subsection{The effect of field discontinuity and large index contrast}

For a bump at waveguide boundaries with a surrounding dielectric, discontinuities in the normal component of  $\mathbf{E}_0$ at the interface makes the product $\delta\boldsymbol \epsilon\cdot\mathbf{E}_0$ ill-defined \cite{johnson2002perturbation}. One strategy to mitigate this issue is to note that if the characteristic lengths of roughness are much smaller than the wavelength, the scattering from this perturbation can be approximated by the long-wavelength limit, for which the quasi-static approximation applies \cite{johnson2005roughness}.

We now restrict our analysis to a waveguide with the permittivity tensor in matrix form given by:
\[
\boldsymbol{\varepsilon} =
\epsilon_0
\begin{pmatrix}
n_e^2 & 0 & 0 \\
0 & n_o^2 & 0 \\
0 & 0 & n_o^2
\end{pmatrix},
\]
where \( n_e \) is the extraordinary index (along \(\hat{x}\)) and \( n_o \) is the ordinary index (along \(\hat{y}\) and \(\hat{z}\)), and with \( n_{\mathrm{clad}} \) for the surrounding material. This corresponds to the LN system we are considering where we use an X-cut TFLN film with waveguides oriented along the the crystal Y axis, so the Z axis is transverse to the waveguide propagation axis. Then using \cite{johnson2005roughness}, the following expressions can be obtained (note that we've redefined the coordinate system so 0 is centered around the sidewall at $x=-d/2$):
\begin{equation}
\label{eq:tangential}
\begin{aligned}
\mathbf{J}_{L}^{\parallel}(y) &= -i\,\omega\,\varepsilon_0\,\frac{1}{2}\left(n_{o}^{2} + n_{\mathrm{clad}}^{2}\right)
\alpha\,E^{\parallel}(x=0,y)\,\delta(x)\,\delta(z), \\
\mathbf{J}_{L}^{\perp}(y) &= -i\,\omega\,\varepsilon_0\,
n_{\mathrm{clad}}^2\,\gamma\,E^{\perp}(x=0^{-},y)\,\delta(x)\,\delta(z).
\end{aligned}
\end{equation}
Here, \( E^{\perp}(x=0^{-}, y) \) represents the field at the interface inside the cladding, while \( E^{\perp}(x=0^{+}, y) \) represents the field at the interface inside the waveguide. The currents \( J_L^{\parallel} \) and \( J_L^{\perp} \) denote the tangential and normal induced currents at the rough interface, respectively. The tangential current can be placed at the interface, whereas the normal current is located at \( x = 0^{-} \), immediately inside the cladding material. The polarizability coefficients \( \alpha \) and \( \gamma \) for a 2D bump can be obtained from
\cite{johnson2005roughness}
 
\begin{equation}
\label{tau}
\alpha(\tau) 
= \frac{2(\tau - 1)}{\tau + 1}
  \left[
    1 + 
    \frac{\tau - 1}{
      \frac{\tau}{\frac{\alpha_{\infty}}{2} - 1} 
      - \frac{1}{\frac{\alpha_{0}}{2} - 1}
    }
  \right],
\quad
\gamma(\tau) 
= \frac{\tau - 1}{\tau}
  \left[
    1 +
    \frac{\tau - 1}{
      \frac{\tau}{\frac{\gamma_{\infty}}{2} - 1} 
      - \frac{1}{\frac{\gamma_{0}}{2} - 1}
    }
  \right].
\end{equation}

Here, the fitted parameters are \(\alpha_{\infty} = 0.8510\), 
\(\alpha_{0} = 3.882\), \(\gamma_{\infty} = 3.905\), and \(\gamma_{0} = 0.7669\). 
We define \(\tau_{o} = \frac{n_{o}^2}{n_{\mathrm{clad}}^2}\) and 
\(\tau_{e} = \frac{n_{e}^2}{n_{\mathrm{clad}}^2}\). 
For the sidewall, \(\gamma(\tau_{e})\) is used for \(E_{x}\) and
\(\alpha(\tau_{o})\) is used for \(E_{y}\) and \(E_{z}\). 
In this formulation, in the limit as \(\tau \to 1\), the standard volume current 
(i.e., \(J \sim \delta \boldsymbol\epsilon\,\mathbf{E}\)) is recovered.

For the upper and lower interfaces, analogous to the sidewall case, the
polarizability matrix for a 3D bump can be obtained from
\cite{johnson2005roughness}.  Here we consider only the tangential
components, with parameters \(\alpha_{\infty} = 1.22\) and
\(\alpha_{0} = 2.85\).

\subsection{Simulation Results of High-Index-Contrast Waveguides: Silicon on Oxide}

\begin{figure}[h]
  \centering
  \includegraphics[width=.65\textwidth]{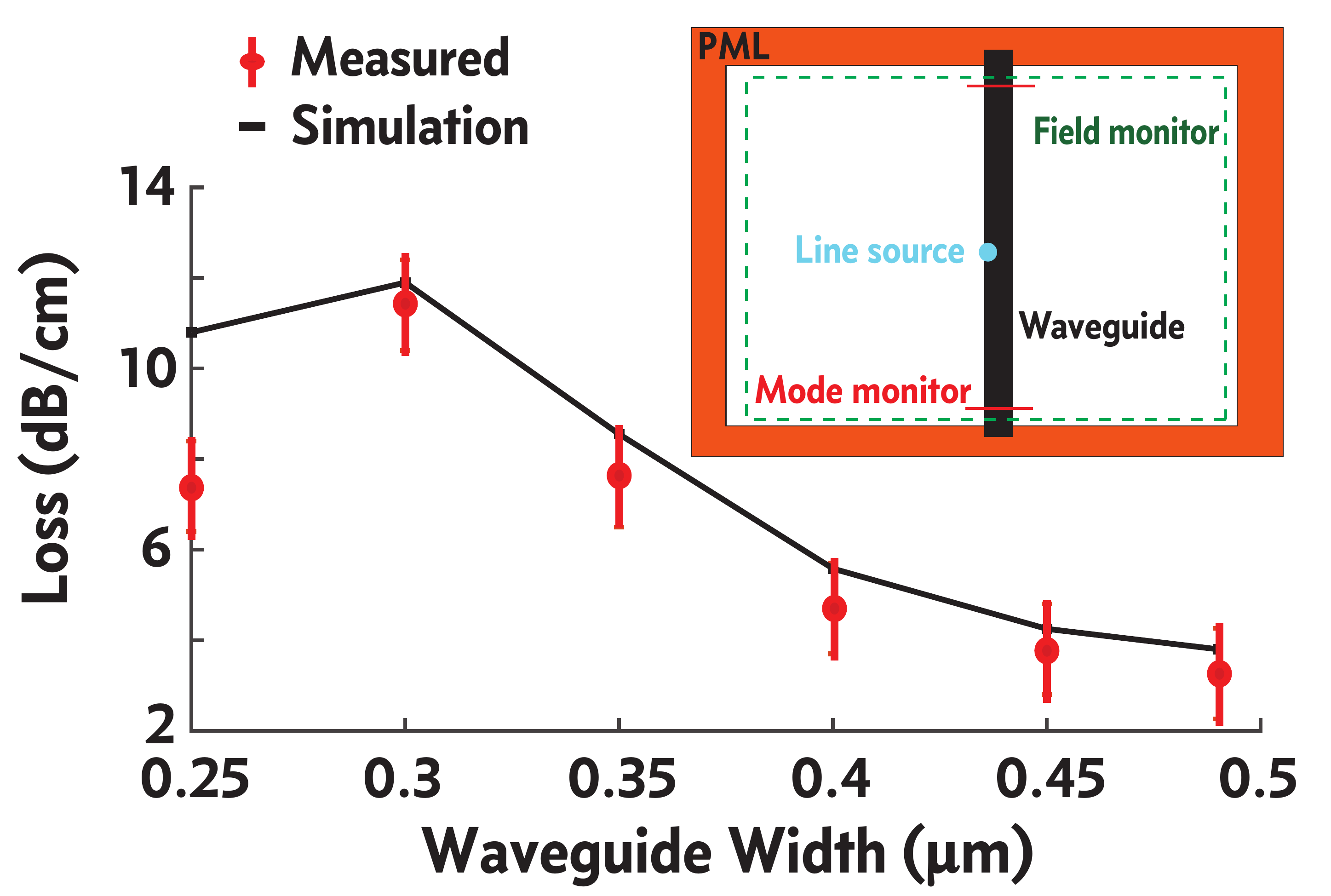}
  \caption{Cladded SOI waveguide simulation with \(H=0.22\,\mu\text{m}\), \(\sigma=2\)~nm, and \(l_c=50\)~nm (adapted from \cite{melati2014unified}). The inset shows the calculation domain in Tidy3D}
  \label{fig:simulationSOI}
\end{figure}

To simulate the radiation profile of the equivalent line source, we used Tidy3D, a GPU-accelerated platform. The calculation domain in each dimension spans approximately \(10\lambda\) to emulate the far field, and the tests confirmed that further increases in the domain size do not alter the results. A field monitor is placed to calculate the radiation intensity in the far field required for \(S_L(\theta,\phi)\) calculation in equation~\eqref{eq:scattering_coefficient}. The line source is placed at the boundary described in the modelling section, and mesh refinement boxes are defined at the boundary to improve the accuracy of the source placement. An important step in the simulation is to place a mode source to subtract the radiation that couples back to the fundamental mode of the waveguide. The custom current source feature of Tidy3D allows for the direct definition of a current source from modal analysis. The setup is illustrated in the inset of Figure~\ref{fig:simulationSOI}. To validate our simulation, we compared our results with experimental data for a silicon-on-oxide waveguide, which features a high index contrast and provides a robust data set for validation. Figure~\ref{fig:simulationSOI} shows the measurement results from \cite{melati2014unified} along with our simulated values.

\subsection{Statistical Parameters of the TFLN Interface}

We used a Bruker \textit{Dimension Icon} AFM equipped with a \textit{ScanAsyst-Air} cantilever operating in PeakForce mode for high‑resolution surface characterization. To obtain representative, artifact‑free statistics, we also fabricated a reference sample of identical chip size and chemical processing but without any patterning.

\begin{figure}[h]
  \centering
  \includegraphics[width=.85\textwidth]{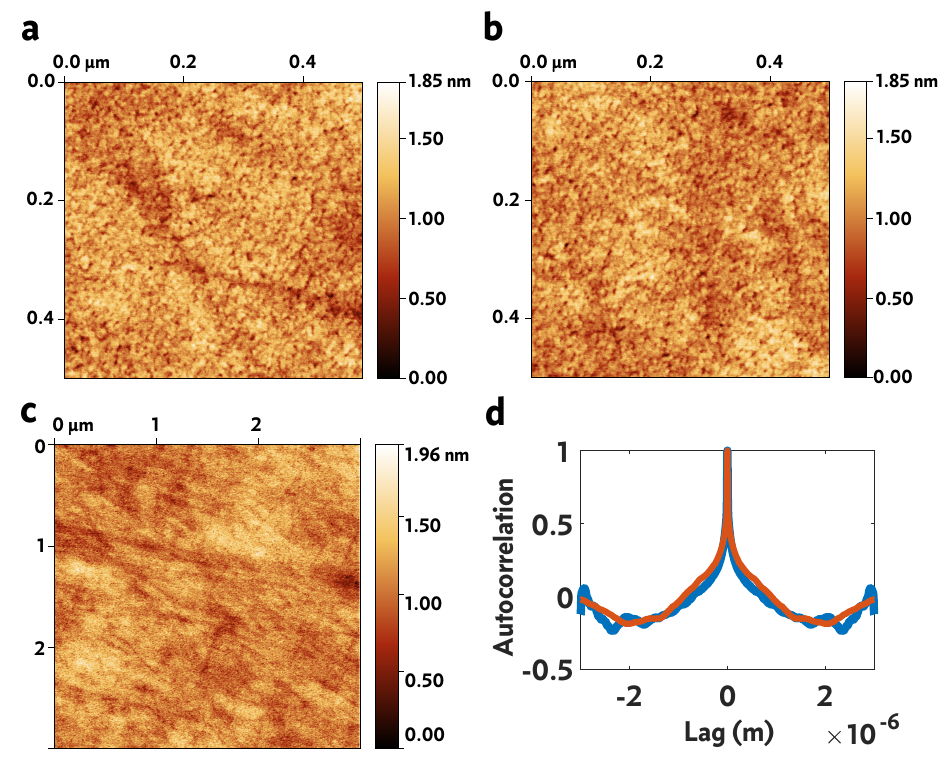}
  \caption{AFM analysis of a CMP‑polished, chemically treated, and annealed TFLN surface. 
    (a) $0.5\times0.5\,\mu\text{m}^2$ scan (1024 lines) showing a smooth surface with polishing trenches. 
    (b) Scan of a different region on the same sample with fewer polishing marks. 
    (c) $3\times3\,\mu\text{m}^2$ scan (1024 lines) used to extract the correlation length. 
    (d) Normalized autocorrelation functions for $3\times3\,\mu\text{m}^2$ (blue) and $5\times5\,\mu\text{m}^2$ (red) areas, both decaying to $1/e$ at $\sim\!80$–$90$~nm. The $5\times5\,\mu\text{m}^2$ scan has $\sim\!2048\times2048$ points.}
  \label{AFM top}
\end{figure}

Figure~\ref{AFM top} presents AFM scans of a sample that was CMP‑polished, followed by HF/Piranha cleaning and annealing. Although this surface is denoted “CMP” in the main text, we note that the post‑fabrication chemical steps dominate the final surface morphology.

Across three locations on the chip, we measured an RMS roughness of $\sim\!180$–$200$~pm and a correlation length of $\sim\!80$–$90$~nm. The data were processed in \textsc{Gwyddion}: mean‑plane subtraction was applied, then median‑based row alignment was used to suppress scan‑line artifacts.

\begin{figure}[h!]
  \centering
  \includegraphics[width=.85\textwidth]{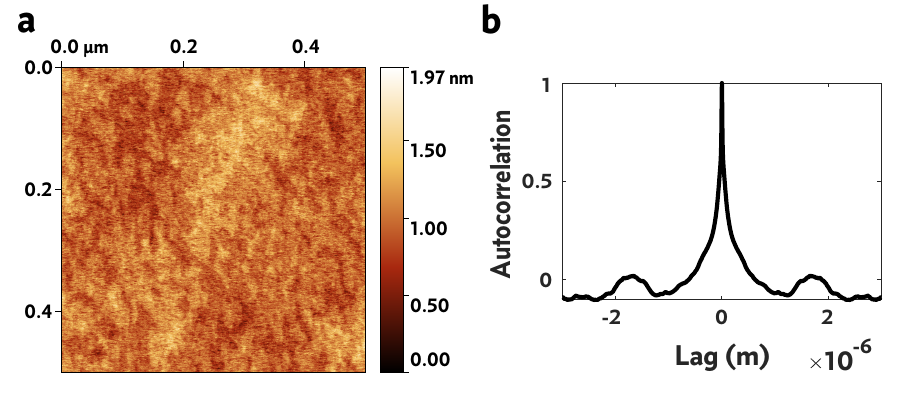}
  \caption{AFM analysis of the argon‑milled, SC‑1‑cleaned TFLN surface.
    (a) $0.5\times0.5\,\mu\text{m}^2$ scan (1024 lines) showing a smooth surface without polishing trenches. 
    (b) Normalized autocorrelation function from a $5\times5\,\mu\text{m}^2$ scan, decaying to $1/e$ at $\sim\!100$ nm.}
  \label{AFM Ar-top}
\end{figure}

Figure~\ref{AFM Ar-top} shows AFM scans acquired with the same instrument and settings for a sample that underwent neutral‑argon‑atom bombardment for trimming, followed by an SC‑1 clean to remove argon implantation from the top surface. The surface exhibits a different morphology, with no polishing trenches; however, the RMS roughness increases to $\sim\!200$–$220$~pm. The correlation length remains $\sim\!100$~nm, although some long‑range features are visible in the scan.

Although we have made every effort to obtain reliable AFM data, we acknowledge that when the surface is extremely smooth and atomic terraces are absent, the estimated roughness and correlation length carry significant uncertainty. In our experience, the choice of leveling procedure can also introduce an additional 10–30~pm variation in these parameters. To account for this, we adopt conservative bounds for the TFLN statistical parameters, namely $\sigma = 0.20$–$0.25\,\text{nm}$ and $L_c = 100$–$150\,\text{nm}$. We also acknowledge that our assumption of isotropic roughness is approximate.

In the following section we estimate the $Q$ factor using two representative lower‑bound cases: (i) $\sigma = 0.2\,\text{nm}$ with $L_c = 100\,\text{nm}$ and (ii) $\sigma = 0.25\,\text{nm}$ with $L_c = 150\,\text{nm}$. These same bounds are used in the main‑text evaluation of interface roughness.

It is worth noting that the optimized Sidewall cleaning protocol developed in this work was recently employed in Ref.~\cite{gruenke2024surface}, where—under ideal annealing conditions—atomic‑terrace formation was demonstrated, suggesting that surface roughness below 0.2 nm is attainable. Further investigation in this direction is left for future work.

\subsection{Simulation Results of High-Index-Contrast Waveguides: TFLN}

By neglecting bending and coupling loss, the total loss in the racetrack resonators can be decomposed into a sidewall-scattering term and a constant term that accounts for the top- and bottom-interface losses together with the intrinsic absorption of lithium niobate:

\begin{equation}
\frac{1}{Q} = \frac{1}{Q_{\mathrm{sidewall}}} + \frac{1}{C},
\label{eq:Q_decomposition}
\end{equation}

where \(1/C\) represents the combined interface and intrinsic material losses. Because the interface contribution is only weakly width-dependent for wide waveguides, we take \(1/C\) from simulations performed on a waveguide with \(w = 3.5\,\mu\mathrm{m}\).

The ridge waveguide considered here has height \(H = 300~\mathrm{nm}\) and slab thickness \(h = 200~\mathrm{nm}\). Absorption loss values are taken from \cite{shams2022reduced}. As discussed in the AFM section of this supplemental material, we used two bounding cases for the top interface statistical parameters: (i) \(\sigma = 0.20\,\text{nm}\), \(L_c = 100\,\text{nm}\), and (ii) \(\sigma = 0.25\,\text{nm}\), \(L_c = 150\,\text{nm}\).

Since we do not have access to the bottom interface, we assume that the bottom surface has the statistical parameters, \(\sigma = 0.20\,\text{nm}\) and \(L_c = 100\text{–}150\,\text{nm}\). This choice is arbitrary but consistent with our CMP-polished TFLN roughness. In addition, we assumed there was no loss of ion implantationation at the buried oxide and lithium niobate interface.

Based on these assumptions, we estimate:
\[
Q_{\text{top}} \approx (90\text{–}200)\times 10^6, \qquad Q_{\text{bottom}} \approx (160\text{–}236)\times 10^6,
\]
Assuming \(Q_{\text{absorption}}\) of about \(160 \times 10^6\), this yields:
\[
\frac{1}{C} \approx (42\text{–}64) \times 10^{6}.
\]

\begin{figure}[h!]
  \centering
  \includegraphics[width=1\textwidth]{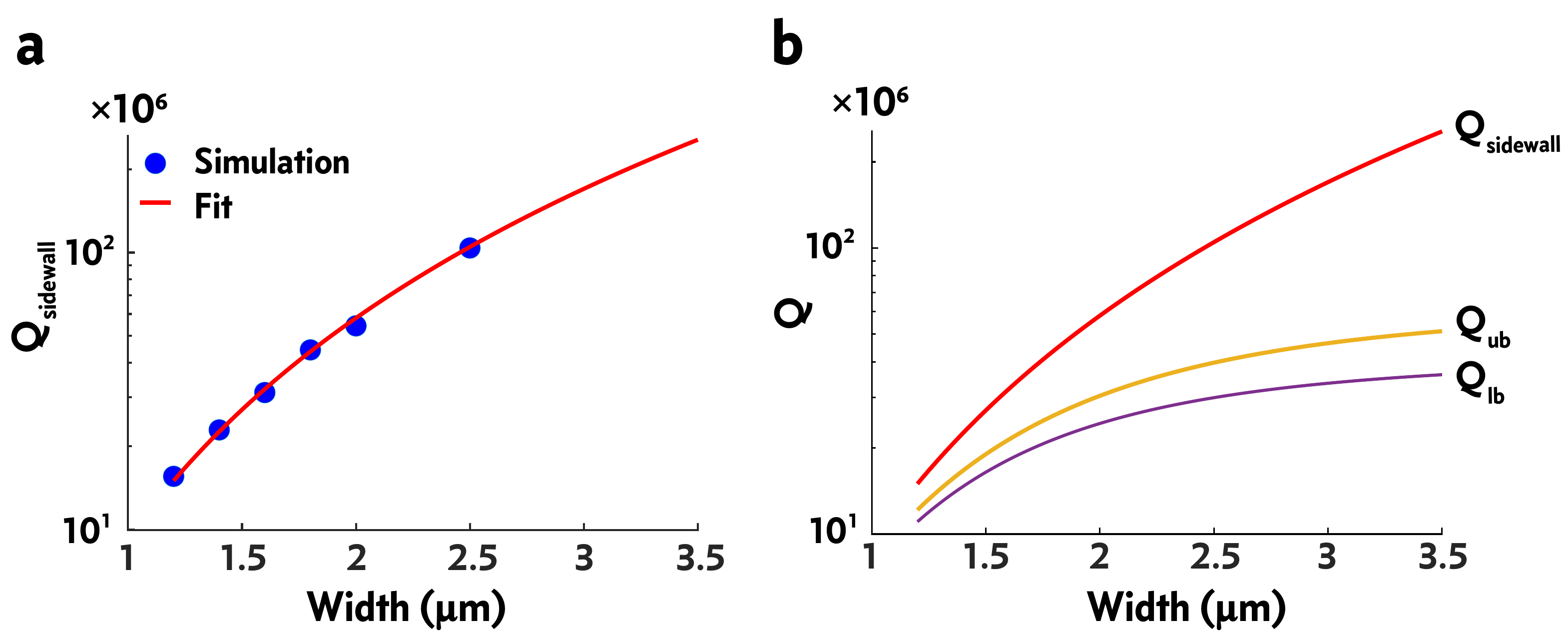}
  \caption{Simulated results for a TFLN waveguide with \(H = 300\,\text{nm}\) and \(h = 200\,\text{nm}\). (a) Simulated \(Q\) from sidewall contribution along with a power-law fit for $L_c = 50$ nm and $\sigma = 1$ nm. The fitted equation is \(Q_{\text{sidewall}} = 15 \left(w / 1.2\,\mu\mathrm{m} \right)^{2.65} \left(1\,\text{nm} / \sigma \right)^2 \times 10^6\). (b) Sidewall contribution (dashed line) and two total quality factor bounds (solid lines) including the proposed range for \(1/C\).}
  \label{TFLNsim}
\end{figure}

From Figure~\ref{TFLNsim}, we can determine the sidewall contribution for devices with \(w = 3.5\,\mu\mathrm{m}\) and \(\sigma = 1\,\text{nm}\) to be \(256 \times 10^6\). For the average roughness measured by AFM of \(\sigma \approx 1.7\,\text{nm}\), we can scale the sidewall contribution to find \(Q_{\text{sidewall}} \approx 89 \times 10^6\).

Using the bound for \(1/C\), the simulated total quality factor is then in the range:
\[
Q_{\text{total}} \approx (29\text{–}37) \times 10^6,
\]
which is consistent with our measured data.

The asymptotic behavior of our measured quality factors around \(Q \approx 29 \times 10^6\) is also consistent with the results reported in~\cite{zhu2024twenty}, which used similar TFLN material provided by the same manufacturing company, NanoLN. In addition, as derived from the main text, we ignore the cross-terms for the top and bottom interfaces, meaning that we underestimate the loss contribution. Therefore, we still believe our suggested bound is an upper limit for finite-width waveguides.

For devices \(w = 1.2\,\mu\mathrm{m}\), the roughness of the sidewall is about \(1.1\,\mathrm{nm}\), yielding an estimated \(Q_{\text{sidewall}} \approx 14 \times 10^6\), which results in total \(Q\) values of \(\approx 11\text{–}12 \times 10^6\) that are also consistent with the measured data.

\subsection{Fano resonance in single side-coupled resonators}
\label{sec:fano}

We measure the transmission spectrum of racetrack resonators by coupling light into and out of the chip through a feedline. The feedline features grating couplers at both the input and output, designed for 1550\,nm, and is side-coupled to the racetrack resonator. Sweeping the tunable Santec TSL-570 diode laser allows us to probe a full spectrum of various modes.

To extract the intrinsic quality factor from this measurement, we model the system as a single-mode cavity with resonance frequency \(\omega_0\), intrinsic loss rate \(\kappa_i\), and external coupling rate \(\kappa_e\). Using the standard coupled mode theory:
\begin{equation}
\frac{d{a}(t)}{dt}
= -\left(i\omega_0 + \frac{\kappa_e+\kappa_i}{2}\right){a}(t)
  + \sqrt{\kappa_e}\, s_{\text{in}}(t),
\label{eq:heisenberg}
\end{equation}
where \(s_{\text{in}}(t)\) is the input field. The output field is given by
\begin{equation}
s_{\text{out}}(t) = s_{\text{in}}(t) - \sqrt{\kappa_e}\,{a}(t).
\label{eq:input_output}
\end{equation}
Assuming a harmonic time dependence \(\sim e^{-i\omega t}\), the cavity amplitude in the frequency domain is given by
\begin{equation}
{a}(\omega) 
= \frac{\sqrt{\kappa_e}}{i(\omega-\omega_0) + \frac{\kappa_e+\kappa_i}{2}}\,
  s_{\text{in}}(\omega).
\label{eq:cavity_amplitude}
\end{equation}

Substituting this into the input--output relation leads to the transfer function:
\begin{equation}
{H}(\omega) 
= 1 - \frac{\kappa_e}{
         i(\omega-\omega_0) + \frac{\kappa_e+\kappa_i}{2}
       }.
\label{eq:transfer_function}
\end{equation}
For quality-factor calculations, it is often more convenient to express the square modulus of the transmission:
\begin{equation}
|{H}(\omega)|^2 
= \text{bg}\,\left|
   1 - \frac{Q\,Q_e^{-1}}{
            1 + 2iQ\,\frac{\omega-\omega_0}{\omega_0}
         }
  \right|^2,
\label{eq:transmission}
\end{equation}
where \(\text{bg}\) can be expressed as a first-order polynomial and represent the background power transmission . The intrinsic quality factor is then defined via 
\[
\frac{1}{Q_i} = \frac{1}{Q} - \frac{1}{Q_e}.
\]

However, to accurately determine the propagation loss in our optical system, it is crucial to account for feedline loss and laser jitter at low scan speeds. Weak reflections in the feedline facets lead to Fabry--Pérot modes that pair to the resonator modes and produce Fano resonances, resulting in asymmetric line shapes~\cite{limonov2017fano}. Traditional symmetric Lorentzian fitting does not capture these asymmetries~\cite{reshef2017extracting}. To include the impact of the Fabry--Pérot response in our fitting, we use a diameter correction method~\cite{khalil2012analysis}, modifying the transfer function as:
\begin{equation}
{H}(\omega) 
= (1+{\varepsilon})
  \left[
    1 - \frac{Q\,\bigl|\hat{Q}_e^{-1}\bigr|\;e^{i\phi}}{
             1 + 2iQ\,\frac{\omega-\omega_0}{\omega_0}
         }
  \right].
\label{eq:fano}
\end{equation}

Here, \( \phi \) is a parameter that accounts for the asymmetry, and \( \varepsilon \) is a complex-valued function with small magnitude that represents the reflection from the feedline facets (analogous to the mismatch between line impedance and load in microwave systems). Since we do not have phase-sensitive measurements for the transmission line in our optical system, we approximate this form by taking its modulus squared.

\begin{equation}
|{H}(\omega)|^2 
= \text{bg}\,\left|
    1 - \frac{Q\,\bigl|\hat{Q}_e^{-1}\bigr|\;e^{i\phi}}{
             1 + 2iQ\,\frac{\omega-\omega_0}{\omega_0}
         }
  \right|^2,
\label{eq:fano_approx}
\end{equation}
where \(\phi\) is a fitting parameter that captures the asymmetry. The intrinsic quality factor is then given by
\begin{equation}
\frac{1}{Q_i}
= \frac{1}{Q}
- \left(\frac{\cos\!\bigl(\phi_0\bigr)}{\bigl|\hat{Q}_e\bigr|}\right).
\label{eq:Qi_fano}
\end{equation}
Here, \( \phi_0 \) is the zeroth-order term in \( \phi \). Another way to interpret the DCM is to rewrite the transfer function as
\begin{equation}
|H(\omega)|^2 = \text{bg}  \left| 1 - e^{i\phi} \cdot \frac{m}{1 + i\,\Omega} \right|^2,
\label{eq:lorentzian}
\end{equation}
where \( m = \dfrac{2\kappa_e}{\kappa_i + \kappa_e} \), \( \kappa_{\text{tot}} = \kappa_e + \kappa_i \), and \( \Omega = \dfrac{2(\omega - \omega_0)}{\kappa_{\text{tot}}} \).

Rearranging yields a Fano-like form:
\begin{equation}
|H(\omega)|^2 = \text{bg} \times \frac{(\Omega - m \sin\phi)^2 + (1 - m \cos\phi)^2}{1 + \Omega^2}.
\label{eq:fano2}
\end{equation}

This expression is equivalent to a Fano shape with a bias term, where the asymmetry is determined by \( m \sin\phi \).

\begin{figure}[htbp]
  \centering
  \includegraphics[width=0.8\textwidth]{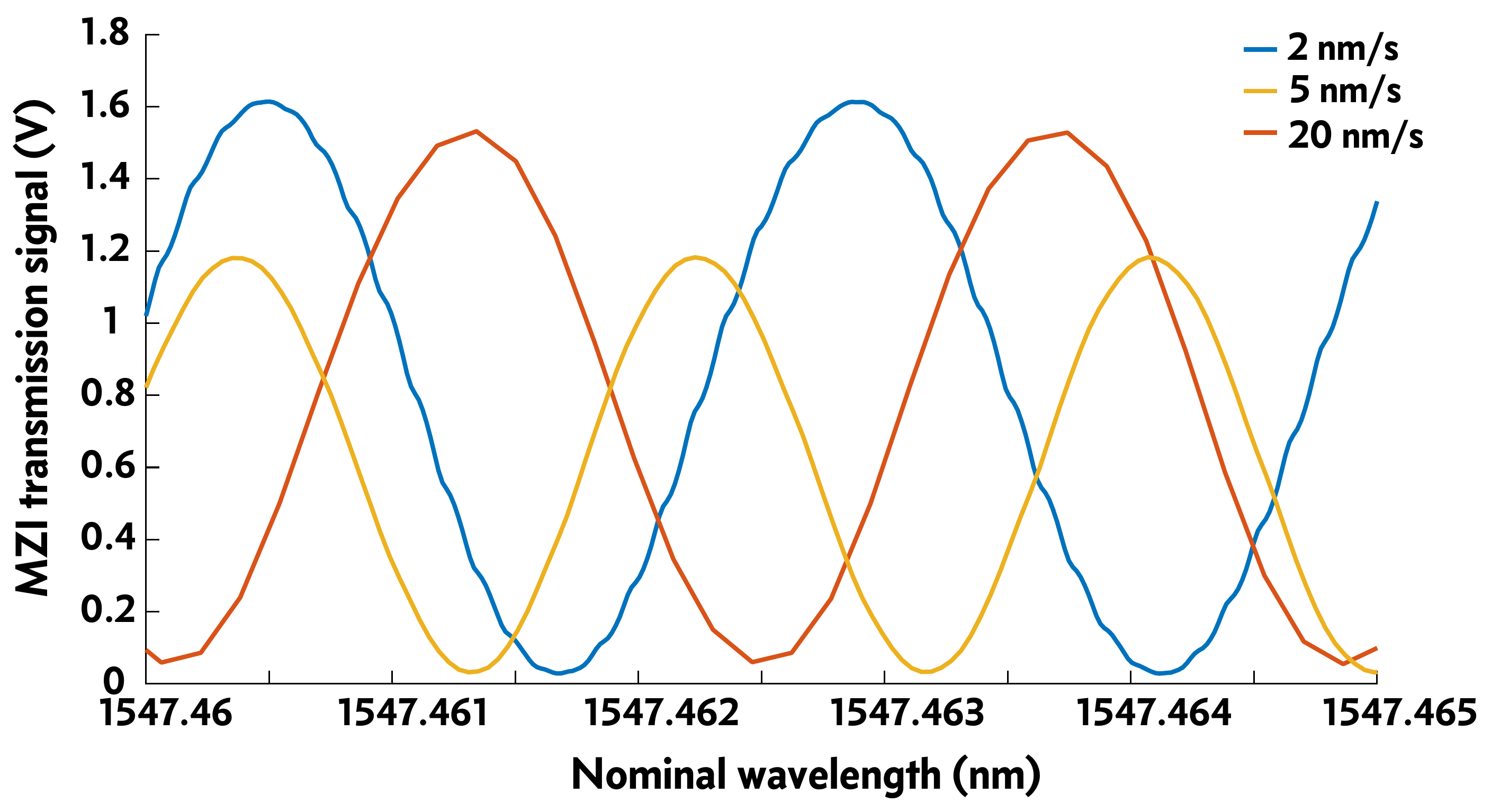}
  \caption{MZI transmission signal for laser wavelength calibration}
  \label{MZI}
\end{figure}

As the device quality factor increases, the measurement becomes more sensitive to the natural jitter of the scanning laser, which degrades at slow scan speeds. The Santec TSL diode laser used in this experiment has a motor to scan the output laser wavelength without any feedback, therefore the actual wavelength output of the laser is not known without any measurement. To characterize the wavelength of the laser, we use a fiber MZI with a known FSR of 325 MHz. The characterization at various scanning rates is shown in Figure~\ref{MZI}. As shown, the laser motor scan speed is not constant, and the jitter becomes more apparent at slower scan speeds below 5\,nm/s. Therefore, all scans are taken at 5\,nm/s at low optical input power.

At these scan speeds, high-\(Q\) resonances may only have a few data points per mode, which is sufficient to extract the linewidth but can affect the fitting near the resonance minimum \(|H(0)|\). Since \(|H(0)|\) directly determines \(Q_i\) from the measured linewidth, interpolation can be used to improve data sampling, though it tends to very slightly decrease the extracted \(Q_i\), ensuring the measured results are conservative estimates.

\begin{figure}[h!]
  \centering
  \includegraphics[width=1\textwidth]{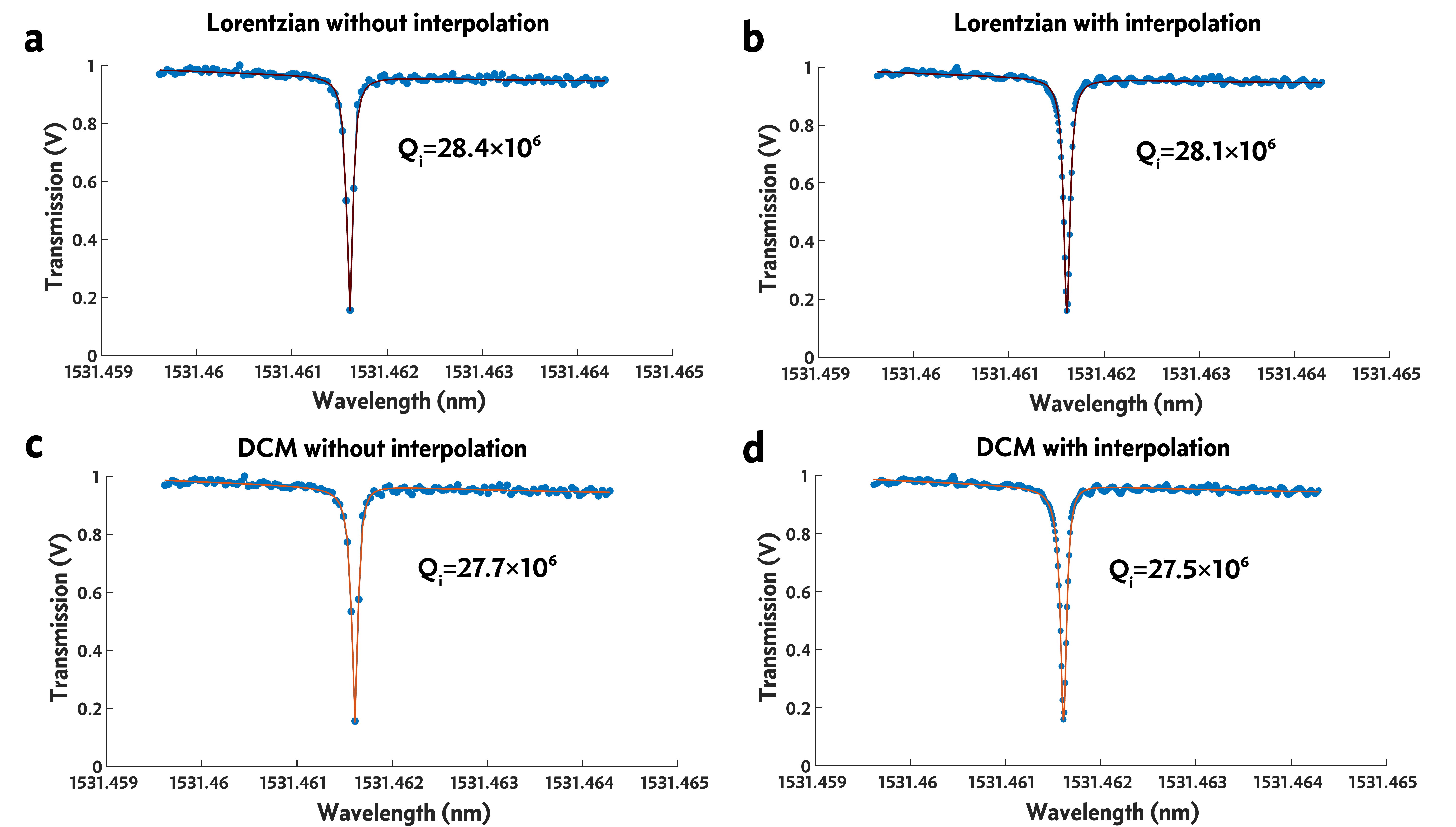}
  \caption{Fitting for Lorentzian and DCM method with and without interpolation.}
  \label{fig:fitting}
\end{figure}

Figure~\ref{fig:fitting} shows the Lorentzian and DCM fits for a representative mode from the main text. As we focus on modes with low asymmetry, the fitted values are expected to be similar.

\subsection{Automatic fitting and symmetry filtering of modes}

\begin{figure}[h!]
    \centering
     \includegraphics[width=0.8\linewidth]{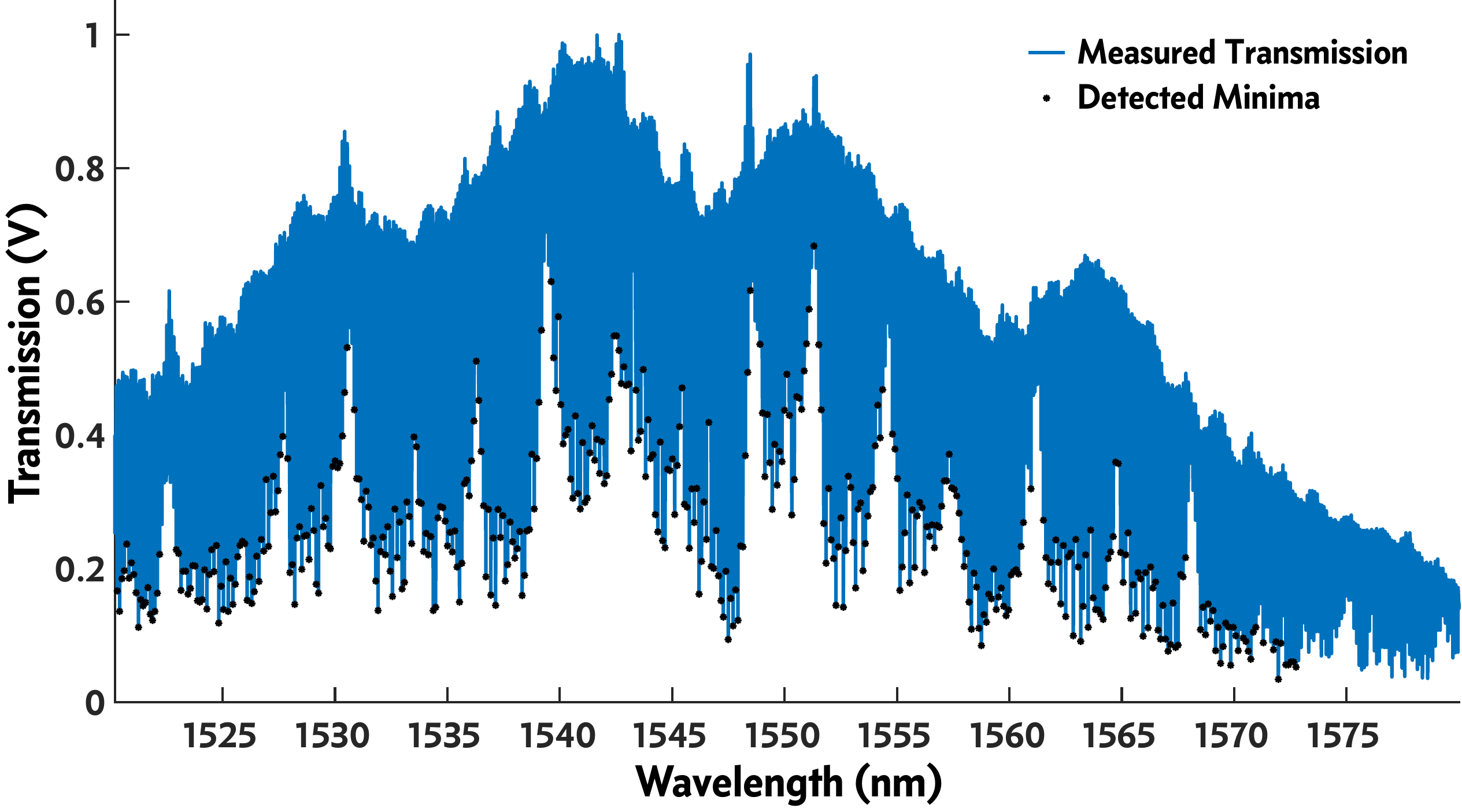}
    \caption{Automatically detected minima.}
    \label{fig:mode_splitting}
\end{figure}

The procedure for mode fitting in the \( w = 1.2\,\mu\mathrm{m} \) device is elaborated in this section. We begin by automatically detecting minima across the full transmission spectrum, as shown in Figure~\ref{fig:mode_splitting}.

\begin{figure}[h!]
    \centering
    \includegraphics[width=0.8\linewidth]{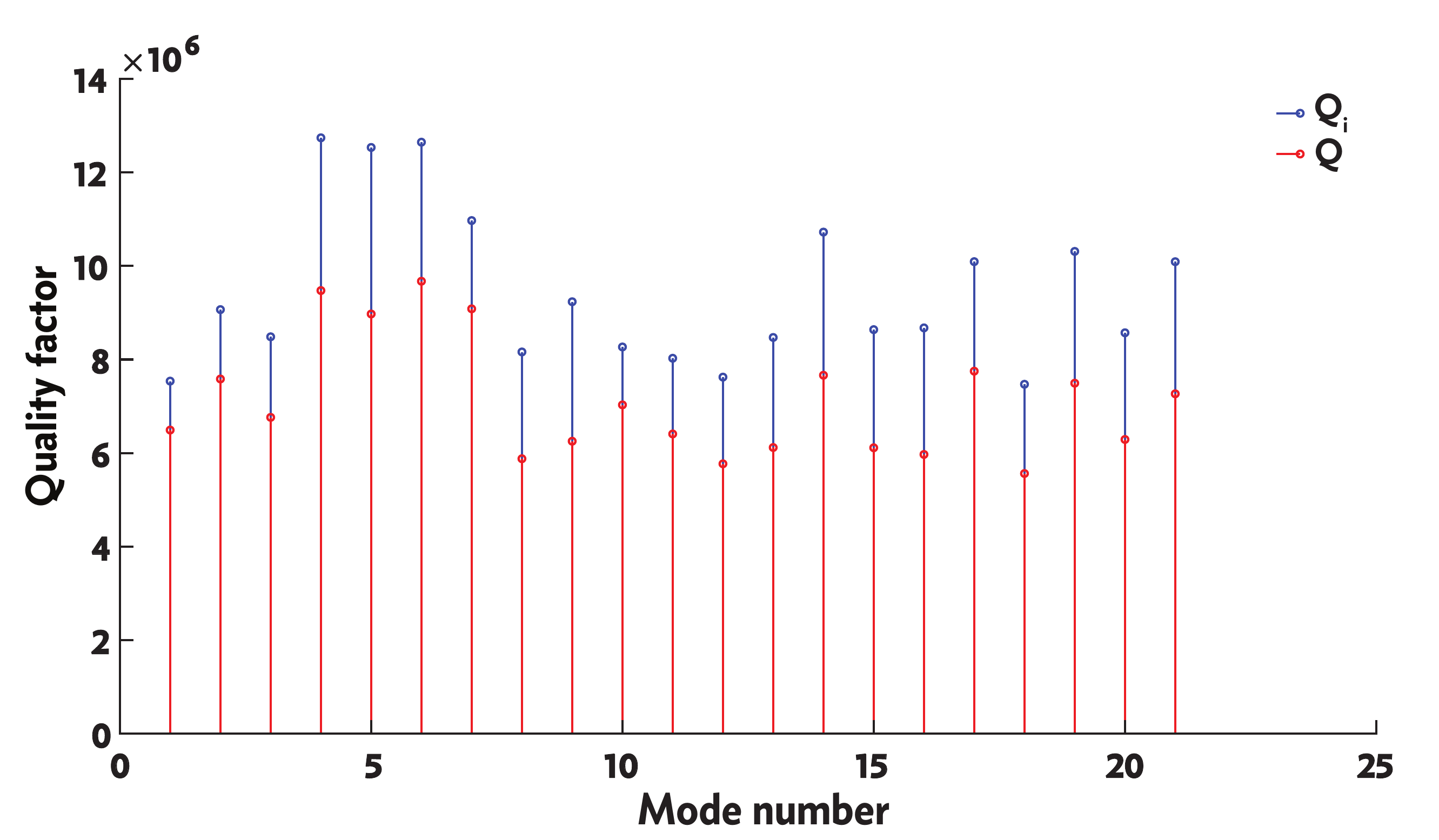}
    \caption{Filtered modes based on fit error and phase asymmetry \( \phi_0 \). Approximately 20 modes satisfy the criteria and are used for statistical analysis.}
    \label{fig:filtered_modes}
\end{figure}

Next, we apply a symmetry-based filtering step by excluding modes with large asymmetry (quantified by the phase parameter \( |\phi_0| > 0.2 \)) and poor fit quality. The result of this filtering is shown in Figure~\ref{fig:filtered_modes}, which includes approximately 20 modes shown in Figure~\ref{fig:filtered_modes}.

\begin{figure}[h!]
    \centering
    \includegraphics[width=1\linewidth]{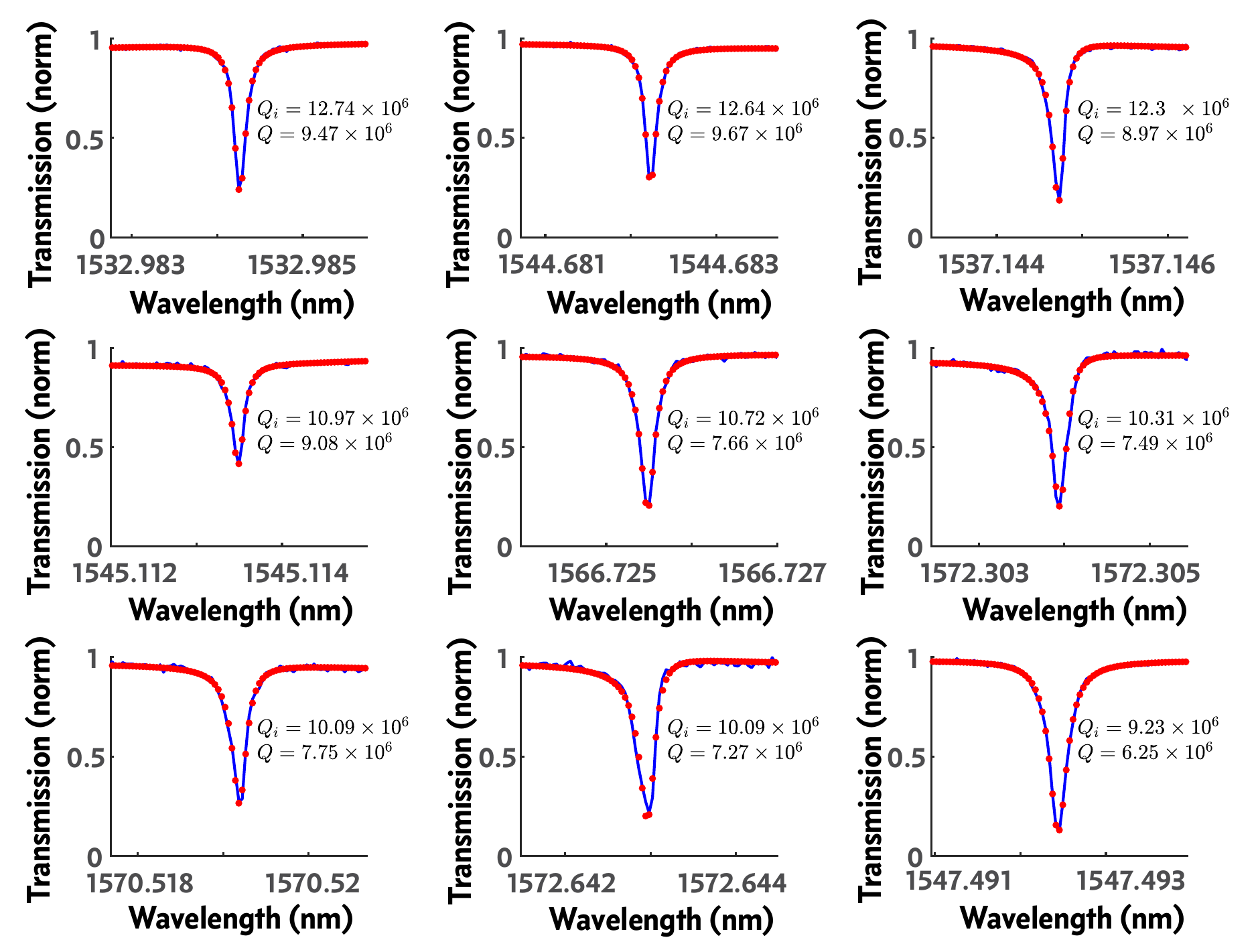}
    \caption{DCM fitting results for the top 9 modes with the highest intrinsic quality factor \( Q_i \) for the \(w = 1.2\,\mu\mathrm{m}\) device. Inset labels show extracted values of \( Q_i \) and loaded \( Q \) for each mode.}
    \label{fig:top9_dcm_fit_w1.2}
\end{figure}

\end{document}